\documentclass[11pt,a4paper]{article}
\usepackage[sort]{cite}
\usepackage{url,colortbl}
\usepackage{xcolor}
\usepackage{bm,amsmath,amssymb}
\usepackage[]{graphicx}
\usepackage[normalem]{ulem}
\usepackage{bbding}
\usepackage{enumitem}
\usepackage{duckuments}
\usepackage{subfig}
\usepackage[margin=2.5cm]{geometry}
\graphicspath{{./Images/}}
\usepackage{authblk}
\usepackage{xr}
\usepackage{hyperref}

\newcommand{\dd}{\text{d}}

\newcommand{\nbri}{T_{\text{bri}}}
\newcommand{\ncon}{T_{\text{con}}}
\newcommand{\nchr}{T_{\text{chr}}}
\newcommand{\nlchr}{T_{\text{l-chr}}}
\newcommand{\nnbs}{T_{\text{nbs}}}
\newcommand{\nlnbs}{T_{\text{l-nbs}}}
\newcommand{\ncnbs}{T_{\text{c-nbs}}}

\begin{document}

\title{Modeling cell differentiation in neuroblastoma: insights into development, malignancy, and treatment relapse}
\author[1]{Simon F. Martina-Perez\thanks{Email: simon.martina-perez@medschool.ox.ac.uk}}
\author[2]{Luke A. Heirene}
\author[3]{Jennifer C. Kasemeier}
\author[3]{Paul M. Kulesa}
\author[2]{Ruth E. Baker}
\affil[1]{School of Medicine and Biomedical Sciences, University of Oxford, UK}
\affil[2]{Mathematical Institute, University of Oxford, UK}
\affil[3]{Children's Mercy Hospital Research Institute, Kansas City, MO, USA}

\date{}
\maketitle
\begin{abstract}
    Neuroblastoma is a paediatric extracranial solid cancer that arises from the developing sympathetic nervous system and is characterised by an abnormal distribution of cell types in tumours compared to healthy infant tissues. In this paper, we propose a new mathematical model of cell differentiation during sympathoadrenal development. By performing Bayesian inference of the model parameters using clinical data from patient samples, we show that the model successfully accounts for the observed differences in cell type heterogeneity among healthy adrenal tissues and four common types of neuroblastomas. Using a phenotypically structured model, we show that alterations in healthy differentiation dynamics are related to cell malignancy, and tumour volume growth. We use this model to analyse the evolution of malignant traits in a tumour. Our findings suggest that normal development dynamics make the embryonic sympathetic nervous system more robust to perturbations and accumulation of malignancies, and that the diversity of differentiation dynamics found in the neuroblastoma subtypes lead to unique risk profiles for neuroblastoma relapse after treatment.
\end{abstract}

\section{Introduction}
Neuroblastoma is a paediatric extracranial solid cancer that arises in the developing sympathetic nervous system \cite{Cohn09, Maris10}. Neuroblastoma tumours, like those of other cancers, are heterogeneous, and contain malignant cells that belong to vestigial or undifferentiated populations in the developmental lineage of healthy patient tissues \cite{Zeineldin22awry, Korber23, Jansky21Origin, Gomez22, Brodeur14Regression}. The heterogeneity of neuroblastoma tumours is well characterised \cite{Gomez22}, and it varies across disease stages, risk classifications, and age groups \cite{Gomez22, Zeineldin22awry, Korber23, Jansky21Origin}. It is known that the cellular composition of patient neuroblastoma samples provides valuable clinical and prognostic insights into the risk associated with the disease \cite{Cohn09, Gomez22}. Additionally, the developmental timing of neuroblastoma initiation plays a critical role in the progression of the disease, as well as the risk of metastasis \cite{Korber23}. A general view, therefore, is that neuroblastoma tumours arise due to errors in differentiation during sympathoadrenal development, leading to an abnormal distribution of cell types in neuroblastoma tumours compared to healthy infant tissues. This cell differentation arrest is thought to mainly affect a sub-population of neuroblasts, as the transcriptome profile of neuroblastoma tumours resembles proliferating neuroblasts or sympathoblasts \cite{Gomez22}. Recently, Jansky \textit{et al.} used single-cell transcriptomic analyses to propose the differentiation trajectories taken by cells in healthy adrenal tissues and neuroblastomas from a common Schwann cell precursor \cite{Jansky21Origin}. In this paper, we use this proposed lineage, together with dynamical systems modelling to relate the altered distributions of cell types in neuroblastoma tumours observed in patient biopsies to possible alterations in the development of foetal sympathoadrenal tissues. 

Dynamic changes in tumour composition provide important insights into possible developmental abnormalities leading to tumour growth, as well as possible treatment strategies. For instance, failure of cells to differentiate from a precursor population to a terminally differentiated cell population is thought to leave a population of multipotent so-called bridge cells \cite{Jansky21Origin, Gomez22}. While this cell type is normally transient, they are observed in high-risk and MYCN-amplified tumours, where they could establish a reservoir of malignant cells to proliferate and develop along different differentiation lineages, potentially leading to therapeutic resistance and relapse \cite{Korber23, Gomez22}. Conversely, tumours comprised of neuroblasts that correctly differentiate along their differentiation pathway may be much more sensitive to treatment. Since immature neuronal progenitors can be re-programmed into benign mature neuronal cells, Brodeur \textit{et al.} have noted that ``[t]he most druggable mechanism is the delayed activation of developmentally programmed cell death regulated by the tropomyosin receptor kinase A pathway. Indeed, targeted therapy aimed at inhibiting neutrophin receptors might be useful in lieu of conventional chemotherapy or radiation in infants with biologically favourable tumours that require treatment'' \cite{Brodeur14Regression}. Therefore, knowing how cell type distributions change during foetal and infant development can provide insights into the key features of pathology and possible therapies.

Knowing how the composition of patient tumours evolves dynamically during neuroblastoma development is of immense clinical relevance, but the available clinical data usually do not allow for this question to be interrogated directly: patient tumour samples are typically only taken at the time of diagnosis or surgical intervention, leading to data sparsity \cite{Schmelz21}. Ordinary differential equations (ODEs) can be used to model differentiation, proliferation, and death in a multi-cell type system as it evolves in time \cite{Bull21}. Since ODE models describe the temporal dynamics of the different tumour components, and are computationally efficient in their implementation, they can reveal the impact of altering key components of the system on the resulting tumour. These components include the ability of cells to convert to other cell types, or their ability to control proliferation \cite{Bull21, Wang23}. ODE models thus provide a means to tease apart possible mechanisms driving cell differentiation and proliferation that might explain the cell and tumour heterogeneity observed across healthy adrenal and neuroblastoma tissue samples. Moreover, there exist well-established computational methods such as approximate Bayesian computation (ABC) and Markov Chain Monte Carlo (MCMC) \cite{Toni:2009:ABC, Toni:2010:SBM, Sunnaker13, Liepe2014AComputation} that can be used to infer system parameters. Such methods can also express the resulting uncertainty in parameter values and model predictions \cite{martinaperez21buq, Sunnaker13, Simpson20}, aiding in the mechanistic, biological, and clinical interpretation of the inferred results. For this reason, ODE models can be used to suggest possible mechanisms of differentiation, proliferation, and death dynamics that might explain clinical data of cell heterogeneity, but cannot be observed directly in practice. 

The structure of this work is as follows. In Section~\ref{section:MainModel} we propose an ODE model of cell differentiation during sympathoadrenal development and show that it can account for the cell type distributions in normal adrenal tissues and four common types of neuroblastoma tumours. In Section~\ref{section:malignancy} we formulate a model for the development of malignancy along cell differentiation trajectories, and show that alterations in healthy differentiation dynamics are related to cell malignancy. In Section~\ref{section:insights} we relate malignancy and differentiation to predict tumour volume growth, and use parameter sensitivity analysis to determine the driving parameters that are involved in tumour regression and growth.

\section{Mathematical model of sympathoadrenal differentiation}
\label{section:MainModel}
\begin{figure}
    \centering
    \includegraphics[width=0.5\linewidth]{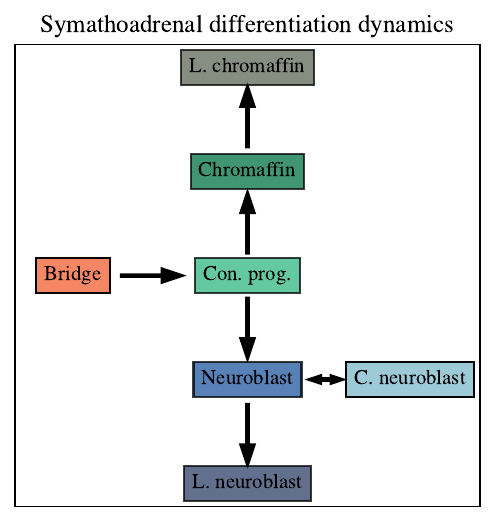}
    \caption{Schematic showing a simplified diagram of differentiation dynamics in the sympathoadrenal lineage as proposed by Jansky \textit{et al.} \cite{Jansky21Origin} using transcriptomic analyses of single cells. Note that the bridge population in this diagram contains both the original bridge population as defined by Jansky \textit{et al.} and the SCP subpopulations.}
    \label{fig:DevelopmentSchematic}
\end{figure}

To formulate an ODE model of cell differentiation during sympathoadrenal development, we follow the proposed cell differentiation lineage proposed by Jansky \textit{et al.}~\cite{Jansky21Origin}. See Figure~\ref{fig:DevelopmentSchematic} for a schematic summarising the developmental dynamics in the model. In the ODE model, we model each cell type as a discrete compartment that is connected to the other compartments representing cell types that are developmentally related. While cellular identity is continuous rather than discrete, we use this framework to relate differentiation dynamics to experimental data collected for different cell types. 

Using single-cell transcriptomic analyses, Jansky \textit{et al.} proposed that the sympathoadrenal lineage from which neuroblastoma cells originate starts with neural crest-derived progenitors termed Schwann cell precursors (SCPs), which either remain in a late precursor or cycling precursor state, or differentiate into a so-called bridge population \cite{Jansky21Origin}. The bridge population then differentiates further into a connecting progenitor population, at which point the lineage bifurcates into a neuroblast branch and a chromaffin cell branch \cite{Jansky21Origin}. Along the neuroblast branch, cells differentiate into neuroblasts, which can terminally differentiate, or remain in a cycling state. Along the chromaffin cell branch, cells differentiate into late chromaffin cells after a transient sojourn in a chromaffin precursor state \cite{Jansky21Origin}. Since the late SCP, cycling SCP, SCP, and bridge populations in the model proposed by Jansky \textit{et al.} only communicate with the rest of the lineage through the link between the bridge population and the connecting progenitor population, we simplify the cell populations by grouping the late SCP, cycling SCP, SCP, and bridge populations into one bridge population which has a proliferative ability. This simplifies the model dynamics and increases interpretability of model outputs. We summarise the resulting model in Figure~\ref{fig:DevelopmentSchematic}.
In this section, we will introduce the ODE model, and calibrate it to experimental data of healthy differentiated tissue in paediatric adrenal medulla tissue to demonstrate that it can successfully capture the trends of cell population make-up. We then use data from different types of neuroblastoma tumours to show that the differentiation dynamics of these patients can be captured by the model. These insights are then used in the next section to relate differentiation dynamics to neuroblastoma initiation, and disease progression.

\subsection{ODE model}
\label{section:ODEModel}
Here, we introduce the main ODE model that will form the basis for the differentiation dynamics in this paper. We denote the total number of cells of type $i$ as $T_i$, such that the conceptual model by Jansky \textit{et al.} \cite{Jansky21Origin}, which is graphically summarised in Figure~\ref{fig:DevelopmentSchematic}, can be described with the following system of coupled ODEs,
\begin{align}
    \dot{T}_{\text{bri}}&=(k_{11} - k_{12}) \nbri \label{eq:cellCountsFirst},\\
    \dot{T}_{\text{con}}&= k_{12}\nbri - (k_{23}+k_{25})\ncon,\\
    \dot{T}_{\text{chr}}&= k_{23}\ncon - k_{34}\nchr,\\
    \dot{T}_{\text{l-chr}}&= k_{34}\nchr - k_{44}\nlchr, \\
    \dot{T}_{\text{nbs}} &= k_{25}\ncon - (k_{57} + k_{56})\nnbs + 2k_{77}\ncnbs,\\
    \dot{T}_{\text{l-nbs}} &= k_{56}\nnbs - k_{66}\nlnbs,\\
    \dot{T}_{\text{c-nbs}} &= k_{57}\nnbs - k_{77}\ncnbs. \label{eq:cellCountsLast}
\end{align}
Here, all transition rates, $k_{ij}$, are assumed to be non-negative. Furthermore, we have assumed exponential growth of the bridge cell population, since this population is known to exhibit a fast and exponential initial expansion during development \cite{Korber23}. The corresponding growth rate is described by the parameter $k_{11}$ in the system. For the neuroblast population, we assume that cycling neuroblasts are a distinct population that yields neuroblasts at rate $k_{77}$. Finally, we assume that late chromaffin and late neuroblast cells die at rates $k_{44}$ and $k_{66}$, respectively. Hence, we consider a linear system of ODEs, in which the ODE for each cell type $i$ may be represented as
\begin{equation}
    \label{eq:TotalEquation}
    \frac{\dd T_i}{\dd t} = \lambda_i T_i + \sum_{j=1}^n k_{ij}T_j.
\end{equation}
Here, we have defined $\lambda_i = k_{ii}$, so that $\mathbf{K} = (k_{ij})_{i,j}$ only contains off-diagonal elements. 

\subsection{Experimental data}
Previously, Jansky \textit{et al.} examined the composition of 17 fresh-frozen healthy human adrenal glands spanning seven developmental time points between Carnegie stage (CS18, seven weeks post-conception), to 17 weeks post-conception \cite{Jansky21Origin}. Cells from these adrenal tissues were analysed using scRNA-seq and assigned cell identities along the sympathoadrenal lineage \cite{Jansky21Origin}. This yielded data about the proportions of the tissues that were made up of each cell type, at the different time points. In addition, samples from human paediatric neuroblastoma patients ($n=498$) were obtained and analysed with the same pipeline. These patients were divided into different categories of the tumour based on their clinical histories, including: tumours with mesenchymal featyures, MYCN-amplified, and TERT/ALT amplified. In contrast to the data for healthy paediatric samples, the data for neuroblastoma patients only contains one time point, which results from a biopsy or tissue sample at the time of surgical resection. Like the data for healthy tissue, these data contain the proportion of the cell population that belongs to the different cell types. Importantly, the data of Jansky \textit{et al.} show that the different neuroblastoma categories can be related to different proportions of cell types, which we will aim to relate to developmental dynamics in the model in the next sections. 

\subsection{ODE for proportions of cell types}
The aim of this section is to introduce a way to relate the observed differences in cell type heterogeneity among healthy tissues and the neuroblastoma tumours that span several disease categories to model outputs. To this end, note that the model introduced in Section~\ref{section:ODEModel} models absolute cell counts, whereas the dataset contains data about the proportions of cells of each type making up the tumour. For that reason, to fit the parameters describing the dynamics of the model in Equations~\eqref{eq:cellCountsFirst}-\eqref{eq:cellCountsLast} one must adapt the model so that it accounts for cell type proportions, rather than absolute cell counts. Denoting the proportion of cell type $i$ in the total population by $R_i$, where $R_i = T_i/N$ and $N = \sum_{i=1}^n T_i$, we find that 
\begin{equation*}
    \frac{\dd R_i}{\dd t} = \frac{N\left(\lambda_i T_i + \sum_{j=1}^n k_{ij}T_j\right) - T_i \sum_{j=1}^n\lambda_j T_j}{N^2},
\end{equation*}
so that
\begin{equation}
    \label{eq:proportionODE}
    \frac{\dd R_i}{\dd t} = \lambda_i R_i + \sum_{j=1}^n k_{ij}R_j - R_i \sum_{i=1}^n \lambda_j R_j.
\end{equation}
Therefore, we obtain a nonlinear system of equations where the unknown parameters are the same as those of the original system of ODEs. That is, the parameters to be estimated from the data are the entries of $\boldsymbol{\lambda}$ and those of $\mathbf{K}$. We discuss the fitting of the parameters to experimental data in Section~\ref{section:MCMC}.

\subsection{MCMC implementation for parameter estimation}
\label{section:MCMC}
Having formulated a model for the evolution of the cell type proportions in Equation~\eqref{eq:proportionODE}, we now aim to calibrate the unknown model parameters to the available data using Bayesian inference. We first note that the interconversion of neuroblasts and cycling neuroblasts in the model as described in Figure~\ref{fig:DevelopmentSchematic} can lead to possible sources of non-identifiability of parameter values, since the effective proliferation rate of the neuroblast population is controlled by the transition rate between neuroblasts and cycling neuroblasts, $k_{57}$, as well as the rate at which cycling neuroblasts proliferate, $k_{77}$. Since the parameter $k_{57}$ effectively controls the amount of time that neuroblasts on average spend in the cycling state, and there is no experimental data to inform this, we set this parameter to unity, leaving only the parameter for the proliferation rate of the cycling neuroblasts to be inferred. That leaves the parameters $k_{11}$, $k_{23}$, $k_{25}$, $k_{34}$, $k_{44}$, $k_{56}$, $k_{77}$ to be identified from the data. 

We use Markov chain Monte Carlo (MCMC) with a standard Haario-Bardenet adaptive covariance and four chains under the assumption of a Gaussian error implemented in Python with the PINTS package~\cite{clerx19pints}. We assume a Gaussian distribution of the error with a standard deviation of $10\%$ in the cell proportions. Using multiple MCMC chains, in this case four,  allows one to assess the ability of the algorithm to converge to the region in parameter space with high posterior probability~\cite{clerx19pints}. As a metric of convergence we use the $\hat{R}$ statistic \cite{clerx19pints}, which is defined as the ratio of the between- and within-chain variances of the chains~\cite{clerx19pints}. The $\hat{R}$ statistic summarises mixing and stationarity, with a value of $\hat{R} = 1$ showing perfect mixing and stationarity of the chains. We use the reference value of $\hat{R} = 1.05$ for four chains \cite{vehtarh21MCMC} and set a maximum number of MCMC iterations to $2\cdot 10^4$. This MCMC procedure yields a posterior distribution over the model parameters, with marginal and bivariate posterior distributions shown in Appendix~\ref{appendix:posterior}. The marginal posterior means are given in Table~\ref{tab:posteriorMeans}. We discuss the interpretation of the posterior distributions in the next section.
  
\begin{table}
    \centering
    \begin{tabular}{c|c|c|c|c|c|c}
    \hline
         $k_{11}$ (wk$^{-1}$)&  $k_{23}$ (wk$^{-1}$)& $k_{25}$ (wk$^{-1}$) &  $k_{34}$ (wk$^{-1}$) & $k_{44}$ (wk$^{-1}$) & $k_{56}$ (wk$^{-1}$) & $k_{77}$ (wk$^{-1}$)\\
         \hline
         1.71$\cdot10^{-3}$ &  1.79$\cdot10^{-3}$ & 3.56$\cdot10^{-3}$  & 7.78$\cdot 10^{-1}$ &  3.36$\cdot 10^{-1}$ & 1.25$\cdot10^{-2}$ & 4.89$\cdot10^{-1}$\\
    \end{tabular}
    \caption{Posterior means arising from MCMC for the parameters in the model given by Equation~\eqref{eq:proportionODE}, fit to data of healthy adrenal medulla samples.}
    \label{tab:posteriorMeans}
\end{table}

\subsection{Cell type distribution in healthy adrenal medulla tissues}
\begin{figure}
    \centering
    \includegraphics[width=\linewidth]{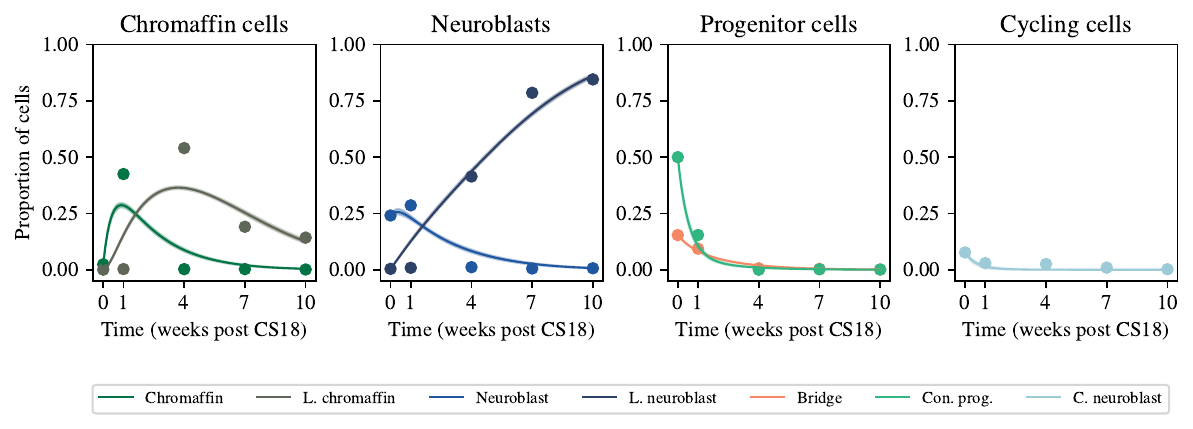}
    \caption{Posterior mean and 95\% posterior confidence interval of predictions for the proportions of different cells types in healthy sympathoadrenal development according to the model in Equation~\eqref{eq:proportionODE} (solid lines) fit to clinical data from Jansky \textit{et al.} \cite{Jansky21Origin} (dots). Simulation parameters are as given in Table~\ref{tab:posteriorMeans}. Note the extremely narrow confidence intervals on the different model prediction traces.}
    \label{fig:NormalDevelopment}
\end{figure}
Having obtained a posterior probability distribution over the unknown model parameters, we simulate the model with the posterior means of this distribution. The resulting model simulations are shown in Figure~\ref{fig:NormalDevelopment}, which shows the posterior mean together with a posterior predictive plot showing the 95\% confidence interval around the posterior mean. These model simulations are in close agreement with the healthy adrenal medulla data and show that the inferred model parameters can recapitulate key features of cell differentiation. These include the population of adrenal medulla tissues by a large proportion of late neuroblasts, while the proportion of progenitor cells, cycling cells, and chromaffin cells all rapidly decrease during development. In particular, the transience of the chromaffin cell population, which occurs in healthy adrenal medulla development but is altered in neuroblastoma \cite{Jansky21Origin} is well-captured by the model. Moreover, the very small proportion of cycling and bridge type cells in the simulation are key features of the model simulations with the posterior mean.

The posterior distributions shown in Appendix~\ref{appendix:posterior} suggest not all parameters in the model are practically identifiable \cite{Komorowski:2011:SRA}. The parameter $k_{25}$ has a broad posterior distribution with a maximum close to zero. This parameter, which we recall describes the transition rate between connecting progenitor cells and neuroblasts, is correlated with the death rate of late chromaffin cells, $k_{44}$, and the proliferation rate of cycling neuroblasts, $k_{77}$. These correlations are intuitive, given that these parameters jointly determine the relative abundances of the different cell types. Therefore, the parameter $k_{25}$ cannot be confidently determined from the cell type proportion data alone in the absence of data regarding the proliferation of individual cell types. Nevertheless, the posterior predictive check in Figure~\ref{fig:NormalDevelopment} shows that the posterior predicted values are distributed tightly around the posterior mean, and form an excellent fit to the experimental data.

\subsection{Cell type distribution in different classes of neuroblastoma}
Knowing that the model in Equation~\eqref{eq:proportionODE} can recapitulate the proportions of cell types in healthy adrenal gland tissues, we wonder if the model can be used to shed light on possible developmental dynamics leading to the proportions of cell types observed in neuroblastoma tumours. Jansky \textit{et al.} \cite{Jansky21Origin} found that the proportion of cell types in neuroblastoma tumours is characteristic for different classes of neuroblastoma. Now, we want to understand which parameter combinations can lead to the different make-up of neuroblastoma in different conditions. Recalling that the data for these tumours is not time-varying, but only contains a snapshot at the time of biopsy or resection, we make the assumption that these cell proportion data represent steady states of the model in Equation~\eqref{eq:proportionODE}. We motivate this by noting that the average age of diagnosis of neuroblastoma is several months after birth, which is on a much longer time scale than the simulations carried out in the previous section, which considered only the 10 weeks post-CS18. Furthermore, to simplify the parameter set and increase parameter identifiability, we also assume that there is no cell death of terminally differentiated cells, \textit{i.e.} we set $k_{44} = 0$.

\begin{figure}
    \centering
    \includegraphics[width=\linewidth]{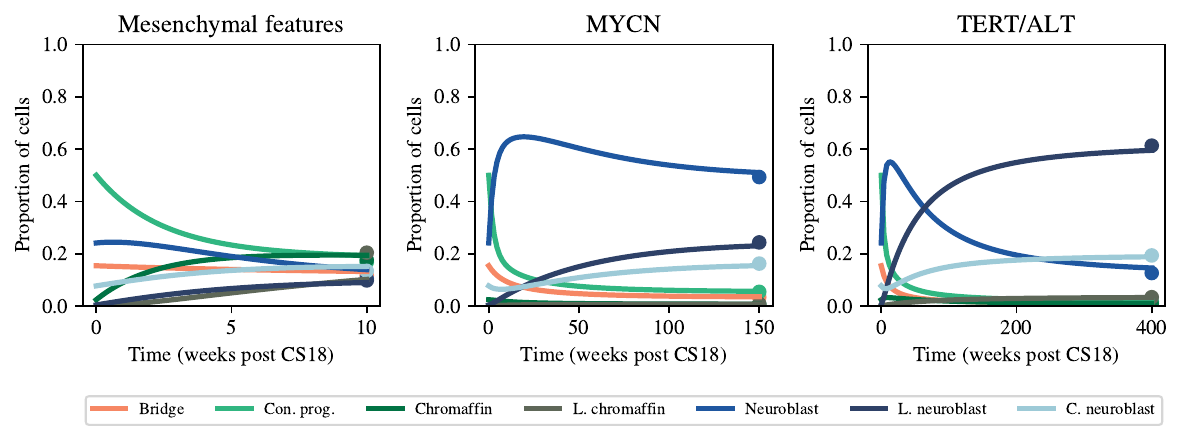}
    \caption{Neuroblastoma categories as steady states of the model in Equation~\eqref{eq:proportionODE}. Model simulations using the parameters in Table~\ref{tab:perturbations} for each of the different neuroblastoma categories.}
    \label{fig:CellIdentitySteadyStates}
\end{figure}

To identify parameter values that give steady states of the model that match the available experimental data, \textit{i.e.}, that yield ${\dd R_i/\dd t = 0}$ for all $i$, we take the cell proportion data for each of the neuroblastoma categories by Jansky \textit{et al.} and substitute them into the right-hand side of Equation~\eqref{eq:proportionODE}. This yields a nonlinear system of equations that cannot be solved algebraically to express the parameter values corresponding to each of the neuroblastoma categories in terms of the available data. Instead, the nonlinear system of equations must be solved numerically to find parameter values that result in the observed data being steady states of the model. We solve for $\dd R_i/\dd t = 0$ numerically using a modified Powell method implemented in the \verb|scipy.optimize| package in Python \cite{virtanen20scipy}. For the initial guess of each of the computational routines, we provide the parameter set of the healthy adrenal medulla as the initial guess for the Powell method. Surprisingly, this yielded a set of parameter combinations that correspond to a steady state of the model in Equation~\eqref{eq:proportionODE} that is equal to each of the cell proportions in the different neuroblastoma categories (see Figure~\ref{fig:CellIdentitySteadyStates}). These parameter values are given in Table~\ref{tab:perturbations}. At this point, we remark that the inference procedure does not guarantee uniqueness of the parameter sets that yield these steady states. 

\begin{table}
    \centering
    \begin{tabular}{l|c|c|c|c|c|c|c}
    \hline
         Category & $k_{11}$&  $k_{23}$& $k_{25}$ &  $k_{34}$ & $k_{56}$ & $k_{57}$ & $k_{77}$ \\
         \hline
         Mes. feat. & 6.58$\cdot10^{-1}$& 1.75$\cdot10^{-1}$& 1.55$\cdot10^{-1}$& 8.96$\cdot10^{-2}$& 7.31$\cdot10^{-2}$& 1$\cdot10^{-1}$& 0\\
         \hline
         MYCN & 6.02$\cdot10^{-1}$&5.08$\cdot10^{-3}$&3.38$\cdot10^{-1}$&2.03$\cdot10^{-2}$&1.00$\cdot10^{-2}$&6.68$\cdot10^{-3}$& 0\\
         \hline
         TERT/ALT &5.86$\cdot10^{-1}$&1.23$\cdot10^{-2}$&2.73$\cdot10^{-1}$&1.97$\cdot10^{-2}$&2.40$\cdot10^{-2}$&7.56$\cdot10^{-3}$& 0
    \end{tabular}
    \caption{All units in (weeks$^{-1}$) Posterior means arising from MCMC for the parameters in the model given by Equation~\eqref{eq:proportionODE}, fit to data of healthy adrenal medulla samples.}
    \label{tab:perturbations}
\end{table}

The parameter values in Table~\ref{tab:perturbations} shows that the different neuroblastoma categories can be distinguished by a few parameters that are noticeably different. While most parameters are of the same order of magnitude to those of healthy tissue, the parameter $k_{23}$, describing the rate of transition between connecting progenitor cells and chromaffin-type cells, spans three different orders of magnitude, being far greater in all disease states than in healthy adrenal medulla development (compare with Table~\ref{tab:posteriorMeans}). This suggests that all disease states are associated with a faster transition to chromaffin-type cells, leading to a large increase of this cell population. Most importantly, though, the bridge cell proliferation rate, $k_{11}$, is much higher in all disease categories than in healthy development, whereas the cycling neuroblast proliferation rate is zero (\textit{versus} 4.89$\cdot10^{-1}$weeks$^{-1}$ for healthy adrenal medulla tissue, see Table~\ref{tab:posteriorMeans}). This fact, together with the fit of zero for the cycling neuroblast proliferation rates, suggests that this model favours bridge cell proliferation and defective subsequent differentiation as a driver for the composition of cell identities within the paediatric tumour samples, rather than maintenance and expansion of the neuroblast population due to neuroblast proliferation in healthy adrenal medulla development.

Having numerically found parameter sets that correspond to steady states of the model that match the data of cell type proportions found in the different neuroblastoma categories, we use these parameters to simulate the ODE model in Equation~\eqref{eq:proportionODE}. For all simulations, and in the absence of information regarding the available initial condition for the neuroblastoma categories, we use the same initial condition as for the healthy adrenal medulla data, since it is believed that at the earliest stages of development, bridge, progenitor and chromaffin cells are most abundant, both in neuroblastoma and in healthy development \cite{Jansky21Origin}. Figure~\ref{fig:CellIdentitySteadyStates} reveals that each of the parameter sets found for the neuroblastoma categories results in ODE solutions that converge to a steady state that matches the experimental data. 

Interestingly, the time taken to reach this steady state is different depending on the neuroblastoma category: the model solutions corresponding to neuroblastoma with mesenchymal features take 10 weeks to reach the steady state (which is in agreement with healthy development), while the solution corresponding to MYCN amplified neuroblastoma takes 150 weeks and that corresponding to TERT/ALT amplified neuroblastoma takes 400 weeks. Since the numerical scheme does not guarantee uniqueness of the parameter sets for the different steady states, these results must be interpreted carefully. However, these results are in roughly in line with the older age of diagnosis of patients with MYCN or TERT/ALT amplification~\cite{Korber23} compared to other neuroblastoma categories. These findings also suggest that these tumours take a longer time to develop and grow, during which genetic defects can accumulate and result in malignant traits, as has been found previously in neuroblastoma~\cite{Korber23}. In the next section we will introduce a model for high-risk clinical features and couple this to the dynamics of development to show that changes in the developmental program of the progenitor cells have a dramatic impact on the development of malignant traits.

\section{Modelling neuroblastoma formation during cell differentiation}
\label{section:malignancy}
Having understood that the heterogeneity in cell types found in healthy adrenal tissue, as well as in the different categories of neuroblastoma tumours, can be related to differences in the differentiation dynamics according to the model, we wish to explore how these dynamics might interplay with the development of malignancy in a developing neuroblastoma tumour. Previously, it has been shown that the time taken for neuroblastoma development as well as the resulting heterogeneity play a role in disease outcomes and progression \cite{Korber23}. We introduce a phenomenological, population structured model to express population malignancy in Section~\ref{section:structuredModel}, which we use to explore the evolution of malignant traits in a tumour in Section~\ref{section:averageMalignancy}. In Section~\ref{section:insights} we use the model to understand the role of developmental dynamics upon disease progression.

\subsection{A phenomenological, population structured model for malignant traits}
\label{section:structuredModel}
During neuroblastoma development, tumour cells randomly acquire a series of mutations before they become malignant \cite{Korber23}. The occurrence of such mutations in cells at early times in the development of the tumour leads various malignant traits to be distributed in the tumour population later on. This distribution can be modelled using a phenotypically structured model \cite{Villa24}. To start constructing such a phenomenological model, we consider malignant trait acquisition by individual cells by assigning each cell a phenotype, $m \in [0,1]$, where $m =0$ represents a completely benign cell and $m=1$ represents a completely malignant cell. Biologically, cells acquire driver mutations stochastically, and can repair damage to DNA. As a result, mathematical models describing the evolution of phenotypic traits of malignant cells are commonly given as stochastic differential equations (SDEs) with a drift, $b \in \mathbb{R}$, which describes a deterministic trajectory leading cells to acquire or lose mutations, and random noise with variance $\sigma > 0$ in the Stratonovich sense, which describes the occurrence of random mutations. By simulating the individual behaviours of a population of cells as they divide, acquire, and lose mutations, one can gain insights into the development of malignant properties in the tumour. Since we are interested in understanding the overall behaviour of the tumour rather than properties of individual cells, we relate how the trajectories of individual cells follow a distribution that evolves over time. For each cell type, $i$, we define the density, $p_i(m,t)$, which gives the density of cells that have malignancy $m$ at time $t$. Conveniently, for an SDE with drift $b$ and variance $\sigma$, and where each cell type $i$ proliferates at rate $\lambda_i \geq 0$, the evolution of $p_i$ for each cell type $i$, is given by a nonlinear Fokker-Planck equation \cite{Risken96},
\begin{equation}
    \label{eq:basicFokkerPlanck}
    \frac{\partial p_i(m,t)}{\partial t} = \lambda_i p_i(m,t) + \frac{\partial}{\partial m}\left(b(m) p_i(m,t)\right) + \frac{\partial}{\partial m}\left(\sigma(m)\frac{\partial p_i(m,t)}{\partial m}\right) + \sum_{j=1}^n k_{ij}  p_{j}(m,t),
\end{equation}
subject to the boundary conditions
\begin{equation}
    \label{eq:basicFokkerPlanckBC}
    \partial_m p_i(0,t) = \partial_m p_i(1,t) = 0,
\end{equation}
for all $t > 0$. Recall that $b(m)$ is a drift term corresponding to the rate at which cells become malignant, and $\sigma(m)$ is a diffusion coefficient that represents the (nonlinear) rate at which cells acquire random driver mutations. As in Equation~\eqref{eq:TotalEquation}, $\lambda_i$ represents the proliferation rate of cell type $i$, and $k_{ij}$ represents the transition rate of cell type $i$ to cell type $j$. 

The PDE description in Equation~\eqref{eq:basicFokkerPlanck} provides a useful framework to keep track of the distribution of phenotypic traits. However, solving the full Fokker-Planck is associated with high computational cost, as well as difficulty in interpreting the results of the population density. Such issues have prompted previous works in phenotypically structured models to reduce Fokker-Planck PDEs to more tractable systems of ODEs \cite{Villa24}. To make progress, and to simplify the following analysis, we further define terms in the model that allow us to reduce the PDE in Equation~\eqref{eq:basicFokkerPlanck} to a system of ODEs that are straightforward to solve numerically. To that end, we observe first that the densities $p_i$ do not define a probability density, because they are not normalised. Rather, as in other works, $p_i$ can be used to express the total number of cells at time $t$, which is given by
\begin{equation}
    \label{eq:TiRedefinition}
    T_i(t) = \int_0^1 p_i(m,t)\dd m.
\end{equation}
Furthermore, we define the functional forms of the drift and mutation rates, $b$ and $\sigma$, so that they are biologically realistic, and can be used to reduce Equation~\eqref{eq:basicFokkerPlanck} to a system of ODEs. To that end, let
\begin{align}
    b(m) & = \gamma (1-m),\\
    \sigma(m) &= \mu m (1-m).
\end{align}
Here, $\gamma$ represents the rate at which cells that have acquired a mutation acquire a new mutation, and the functional dependence of $b$ on $m$ is chosen to reflect the fact that the number of possible new driver mutations decreases as the number of malignant traits in a cell increases. Likewise, $\mu$ represents the rate at which random mutations are acquired, and the functional dependence on the malignancy is chosen to reflect two key characteristics. On the one hand, cells with few malignant traits have intact cell cycle checkpoints and DNA repair mechanisms, implying $\sigma(m) \to 0$ as $m \downarrow 0$. On the other hand, cells that have many malignant traits will preserve their malignant traits and will not lose them through repair, implying $\sigma(m) \to 0$ as $m \uparrow 1$. Having defined the drift and diffusion coefficient, we seek to reduce the PDE in Equation~\eqref{eq:basicFokkerPlanck} to an ODE. We define
\begin{equation}
    M_i(t) = \int_0^1 mp_i(m,t)\dd m.
\end{equation}
In Appendix~\ref{section:MiDerivation}, we show that $M_i$ is governed by the ODE,
\begin{equation}
    \label{eq:MalignantEquation}
    \frac{\dd M_i}{\dd t} = (\lambda_i-\gamma - 2\mu) M_i +(\gamma+\mu)T_i  + \sum_{j=1}^n k_{ij}M_j.
\end{equation}
Moreover, $T_i$ as defined in Equation~\eqref{eq:TiRedefinition} indeed follows Equation~\eqref{eq:TotalEquation}, as can be seen by integrating Equation~\eqref{eq:basicFokkerPlanck} over $m$ and using the boundary conditions in Equation~\eqref{eq:basicFokkerPlanckBC}. Since $T_i$ and $M_i$ can be thought of as the total number of cells of type $i$, and the expected number of cells of type $i$ that are malignant, the reduction of the phenotypically structured model in Equation~\eqref{eq:basicFokkerPlanck} yields a straightforward system of ODEs that can be used to probe cell population dynamics and how they interface with malignant traits across the population.

\subsection{Modelling average malignancy}
\label{section:averageMalignancy}
Having established a way to describe the difference between the total number of cells of each cell type, and the total number of cells weighted by their malignant traits in Equation~\eqref{eq:MalignantEquation}, we use this framework to express to what extent each of the different cell populations exhibits malignant traits. The reason for doing this is that knowing the extent to which a cell population is comprised of malignant cells is highly informative for treatment outcomes and the potential for post-treatment relapse \cite{Gomez22, Schmelz21, Zeineldin22awry}. For example, it has been suggested that the cycling population of bridge cells might be a reservoir of potential malignancy \cite{Jansky21Origin, Zeineldin22awry}. However, it is difficult assess the extent to which any of the cell populations express malignant traits by only looking at cell numbers, $M_i$ and $T_i$. To directly analyse the malignant traits in each of the cell populations, we aim to quantify the average malignancy of the cells of a given cell type. To do this, we define the average malignancy as $\mathcal{M}_i = M_i/T_i \in [0,1]$. The benefit of analysing this quantity is that it is independent of cell count and can be compared between cell types with different cell counts. In Appendix~\ref{appendix:malignancy}, we derive that
\begin{equation}
    \label{eq:malignancy}
    \frac{\dd \mathcal{M}_i}{\dd t} = (\gamma + \mu) - (\gamma + 2\mu)\mathcal{M}_i + \sum_{j=1}^n \frac{R_j}{R_i} k_{ij} (\mathcal{M}_j - \mathcal{M}_i).
\end{equation}

\begin{figure}
    \centering
    \includegraphics[width=\linewidth]{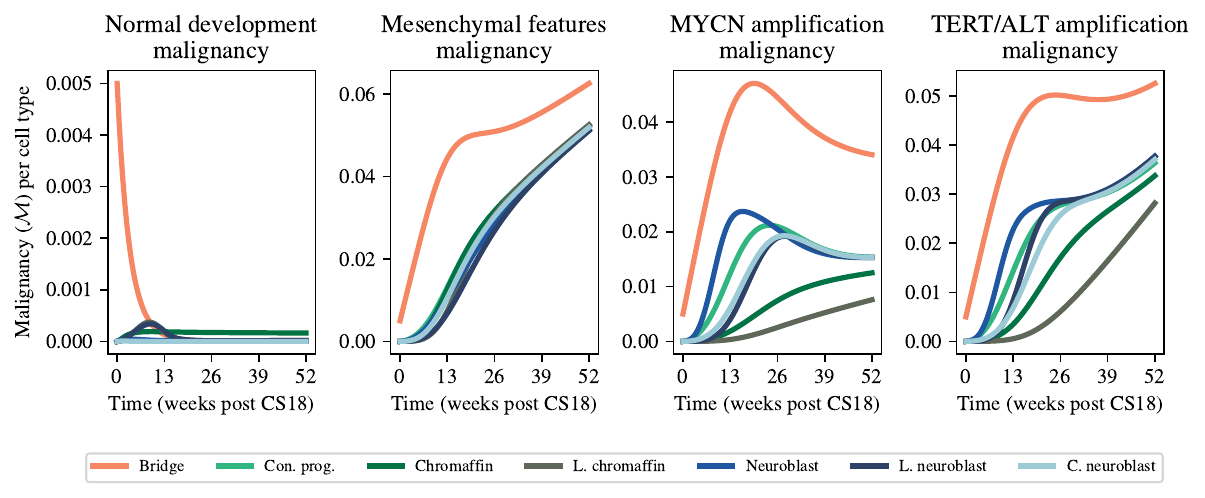}
    \caption{Simulations of average malignancy according to the model in Equation~\eqref{eq:malignancy} coupled with Equation~\eqref{eq:proportionODE}. From left to right, parameters for proliferation rates, $\lambda_i$, and transition rates, $k_{ij}$ belong to different parameter regimes corresponding to healthy development or neuroblastoma disease categories. In all simulations, the initial condition for malignancy is taken to be $\delta = 0.005$, and $\gamma = \mu = 5\cdot10^{-2}\text{weeks}^{-1}$.}
    \label{fig:malignancy}
\end{figure}

Note that this equation only depends on the malignancies and relative abundances, $R_i$, so that this system can be simulated for each of the healthy and neuroblastoma cases found previously, for the simulations in Figure~\ref{fig:NormalDevelopment} and \ref{fig:CellIdentitySteadyStates}, respectively. Using these parameters, and setting $\gamma = \mu = 5\cdot10^{-2}\text{weeks}^{-1}$, we simulate the system defined by Equation~\eqref{eq:malignancy} coupled with Equation~\eqref{eq:proportionODE} for each cell population in each of the different parameter regimes (\textit{i.e.} for normal development, and for each of the different neuroblastoma categories). For the initial condition of the proportions, we set the same proportions as in Figure~\ref{fig:CellIdentitySteadyStates}. For the initial mean malignancy, we use $\mathcal{M}_i = \delta = 0.005$ for all cell types and all simulations. We show the results of these simulations in Figure~\ref{fig:malignancy}. For all conditions we simulate between $t = 0$ weeks post-CS18 and $t = 52$ weeks post-CS18.

Figure~\ref{fig:malignancy} shows several interesting things. Most strikingly, the average malignancy for all cell types during normal development rapidly diminishes and remains stably close to zero for all cell types. In contrast, for all of the neuroblastoma categories, the average malignancy rises rapidly in the first weeks post-CS18. This results in cell populations with appreciable numbers of cells with malignant traits. While our model does not capture this, increasing numbers of cells with malignant traits is generally associated with higher risk of metastasis and cancer progression~\cite{Zeineldin22awry, Cohn09}. Noting that the only difference between the simulations in healthy development and neuroblastoma development in Figure~\ref{fig:malignancy} is the development dynamics, our model findings suggest that not only do healthy developmental dynamics contribute to controlling the proportion of cells with malignant traits within the population, but also that perturbations of these can be associated with the rapid accumulation of malignant traits. 

\section{Insights into neuroblastoma tumour progression}
\label{section:insights}
The findings in the previous section revealed that the average malignancy of the cell populations involved in neuroblastoma is strongly influenced by the rates at which cell differentiation occurs. To further understand under which conditions progression of a benign mass into a tumour is predicted by our model, we seek to further reduce the complexity of the system under investigation by analysing the total tumour volume relative to the cell population. Whereas this only gives a proxy for the relative proportion of cells within the population that are malignant, it does provide a means to investigate under which conditions a benign neoplasia can progress into a malignant tumour. This quantity will then be used to investigate which parameters in the model are most strongly associated with the progression of the neuroblastoma tumour, yielding insights into treatment response and relapse. 

\subsection{Modelling of neuroblastoma tumour volume}
As a proxy for the share of the total tissue mass that is cancerous, we introduce the quantity $V$, which we define as
\begin{equation}
    V = \sum_{i=1}^n V_i = \sum_{i=1}^n R_i \mathcal{M}_i,
\end{equation}
\textit{i.e.}, it is the weighted tumour burden on the patient. Expressing this quantity using the governing equation for the cell proportions, $R_i$, Equation~\eqref{eq:proportionODE}, and that of the average malignancy, $\mathcal{M}_i$, Equation~\eqref{eq:malignancy}, we find that
\begin{equation}
    \label{eq:Veq}
    \frac{\dd V}{\dd t} = (\gamma + \mu) \sum_{i=1}^n \lambda_i R_i - (\gamma + 2\mu)V + \sum_{i=1}^n \lambda_i V_i - V\sum_{i=1}^n \lambda_iR_i,
\end{equation}
subject to  $V(0) = V_0$. Note that Equation~\eqref{eq:Veq} is in the form 
\begin{equation}
    \frac{\dd f}{\dd t} = a(t) - b(t) f(t),
\end{equation}
and therefore admits a solution in the form
\begin{equation}
    V(t) = \exp\left(\int_0^t -b(\xi)\dd\xi\right)\left(\int_0^t\exp\left(\int_0^\zeta b(\xi)\dd\xi\right)a(\zeta) \dd\zeta + V_0\right),
\end{equation}
with 
\begin{align*}
    a &= (\gamma+\mu)\sum_{i=1}^n \lambda_iR_i + \sum_{i=1}^n \lambda_i V_i,\\
    b &= \gamma + 2\mu - \sum_{i=1}^n \lambda_iR_i.
\end{align*}

\begin{figure}[htb]
    \centering
    \includegraphics[width=\linewidth]{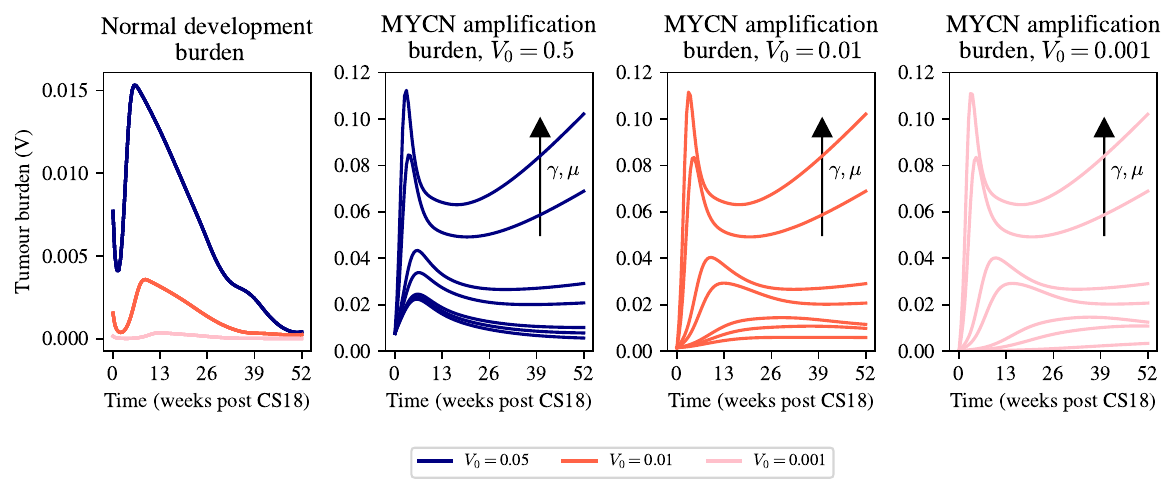}
    \caption{Tumour burden dynamics associated with model simulations according to Equation~\eqref{eq:Veq} parameterised according to healthy conditions (left-most panel) and MYCN-amplified neuroblastoma. Healthy parameter values are as given in Table~\ref{tab:posteriorMeans}, and MYCN-amplified parameter values as given in Table~\ref{tab:perturbations}. The parameters $\mu$ and $\gamma$ are varied across simulations as described in the text.}
    \label{fig:tumourBurden}
\end{figure}

As an initial investigation of neuroblastoma tumour progression in the different parameter regimes studied in this paper, we simulate the tumour burden in the case of healthy development dynamics, and in the case of MYCN amplified neuroblastoma, as various of the system parameters are studied. For the healthy development dynamics we again take the posterior means identified in Table~\ref{tab:posteriorMeans}, and for the MYCN amplified neuroblastoma we take the parameters given in Table~\ref{tab:perturbations}. We then simulate for a varying range of initial conditions: $V_0 = 0.05, 0.01, 0.001$. We also use different values for the mutation parameters $\mu, \gamma$: keeping $\mu = \gamma$, we take the values $1\cdot10^{-1}$, $5\cdot10^{-2}$, $1\cdot10^{-2}$, $5\cdot10^{-3}$, $1\cdot10^{-3}$, $5\cdot10^{-4}$, and $1\cdot10^{-4}\text{weeks}^{-1}$. The different simulation traces are given in Figure~\ref{fig:tumourBurden}.

The plots in Figure~\ref{fig:tumourBurden} demonstrate that the tumour burden dynamics as modelled by Equation~\eqref{eq:Veq} vary greatly between healthy and neuroblastoma conditions, but also within these conditions as the critical parameters $\mu$ and $\gamma$ are varied. In the case of healthy development, reminiscent of the malignancy profiles in Figure~\ref{fig:malignancy}, the tumour burden very rapidly shrinks, across all initial conditions and values of $\mu$ and $\gamma$. On the other hand, in the case of MYCN-amplified neuroblastoma there are clear differences between the different parameter regimes for $\gamma$ and $\mu$. When $\gamma$ and $\mu$ are small enough, the tumour burden remains constant or decreases in a manner similar to that of healthy development, whereas when the parameters are increased, the tumour burden in MYCN amplified neuroblastoma rapidly increases and escapes. 

Put together, our model simulations suggest that whereas healthy development dynamics can provide protection against tumour burden escape across a wide range of initial conditions and rates of mutation, this is no longer the case when these dynamics are altered. For that reason, it is important to ascertain which model parameters have the most influence on tumour burden to understand under which conditions one might expect tumour progression, and relapse after treatment.

\subsection{Global parameter sensitivity analysis}
So far, we have observed that varying the parameters of malignancy progression, $\mu$ and $\gamma$, and those describing cell differentiation, influences whether the tumour will be eliminated or grow. In what follows, we would like to understand which parameters most influence the tumour burden. Given that the parameter space contains nine free parameters, a combinatorial grid search is not feasible to explore the parameter space. Parameter sensitivity analysis is a valuable tool for identifying the parameters to which a model's output of interest is most sensitive. One widely used and effective method for global parameter sensitivity analysis is the calculation of Sobol indices, which quantify the contributions to the variance of a scalar model output from different model parameters \cite{Sobol′2001}. In this section, we will use Sobol indices to establish the key model parameters influencing total tumour burden.

\begin{figure}[htb]
    \centering
    \includegraphics[width=\linewidth]{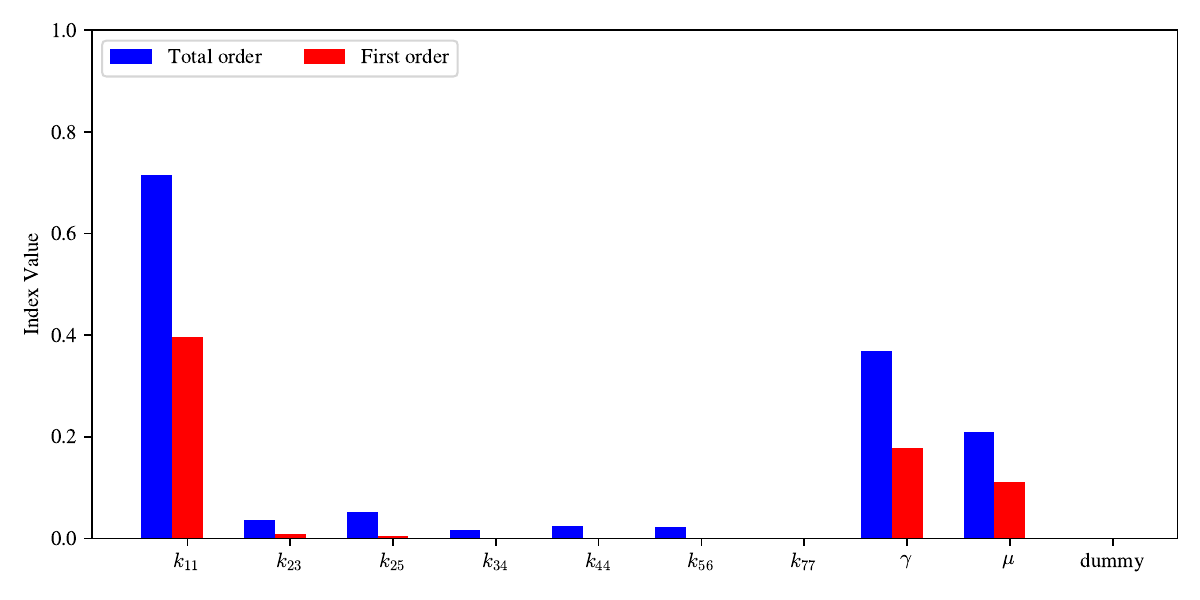}
    \caption{Total and first order Sobol sensitivity indices (in blue and red, respectively) for the tumour burden at week 52. The parameter bounds were set as $10^{-3}-1$ for all parameters in Table \ref{tab:posteriorMeans} except for $k_{77}$ for which they were set at $10^{-19}-10^{-17}$. The bounds for $\mu$ and $\gamma$ were set as $10^{-4}-10^{-1}$. A dummy parameter was added to the analysis to estimate uncertainty within the sensitivity indices as in \cite{Marino2008}.}
    \label{fig:ParameterSensitivity}
\end{figure}

To calculate the Sobol indices, we first define bounds for the model parameters and then sample using a quasi-random sequence as detailed in \cite{Homma1996, Sobol′2001} to generate multiple parameter sets within these bounds. The parameters included in this sensitivity analysis are the same as those listed in Table \ref{tab:posteriorMeans}, along with mutation parameters $\mu$ and $\gamma$. For each parameter set, the tumour volume at week 52 was computed using Equation (\ref{eq:Veq}). First and total order Sobol indices were then calculated using the \verb|Salib| package \cite{Herman2017}, and the results are shown in Figure \ref{fig:ParameterSensitivity}. The sensitivity analysis reveals that the parameters with the greatest influence on tumour burden at week 52 are the growth rate of the bridge cell population, $k_{11}$, as well as the mutation parameters $\mu$ and $\gamma$. This is in agreement with the idea that the bridge population is a reservoir for malignancy, and that its continued expansion after a normal development window might drive the development of neuroblastoma. 
\label{section:paramSensitivity}

\subsection{Conditions for spontaneous regression}
\begin{figure}
    \centering
    \includegraphics[width=0.8\linewidth]{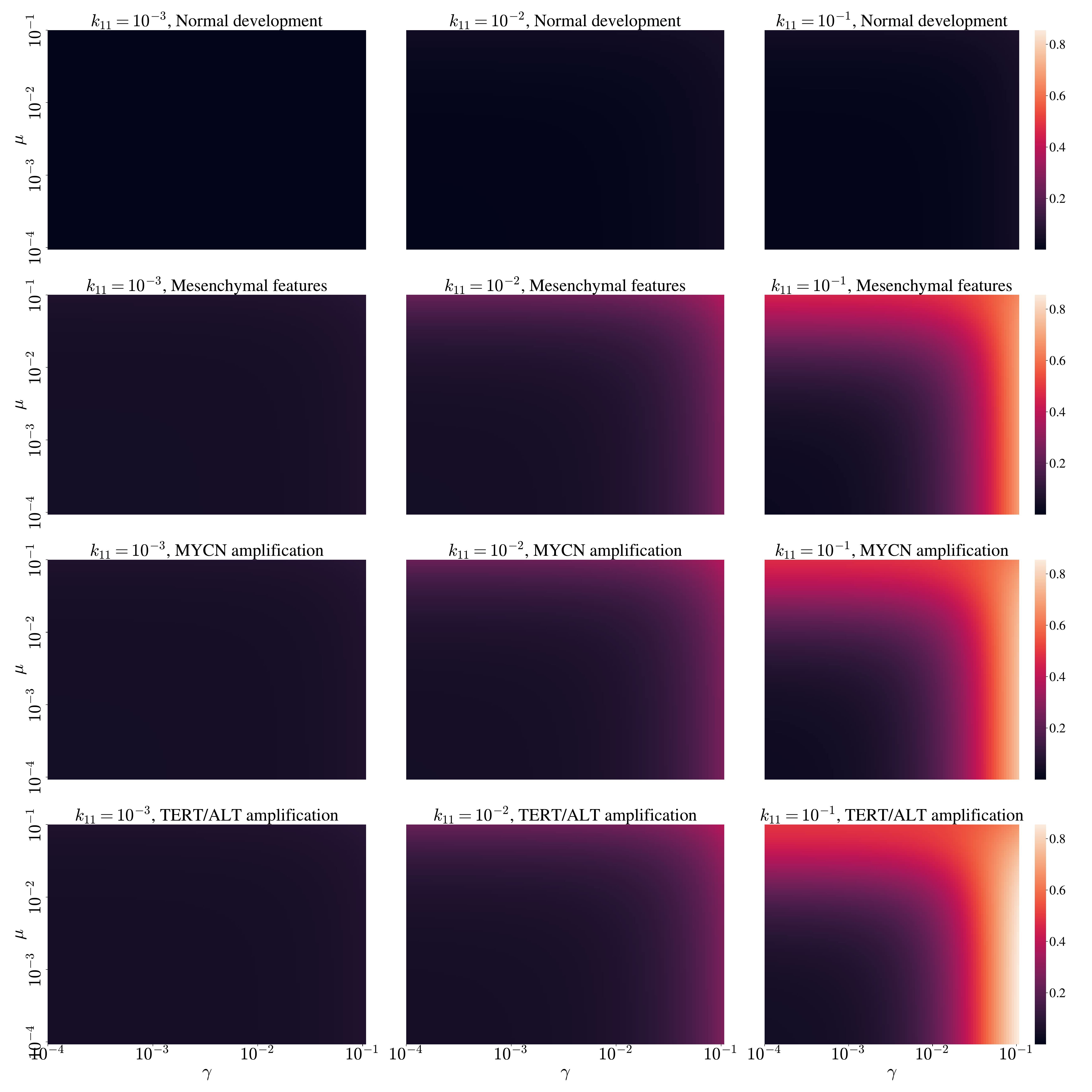}
    \captionsetup{width=1\textwidth}
    \caption{Plots of the tumour burden at week 52 as $\gamma$ and $\mu$ vary at different values of $k_{11}$. All other parameters were set dependent on separate situations corresponding to either normal development or a tumour type with mesenchymal features, MYCN amplification orTERT/ALT amplification.}
    \label{fig: psweeps}
\end{figure}

Having identified that the mutation rates, $\mu$ and $\gamma$, and the bridge cell proliferation rate, $k_{11}$, are the three key parameters that most influence tumour burden at week 52, we can further explore the role of these three parameters in tumour escape or regression. As observed in Figure~\ref{fig:tumourBurden}, $\mu$, $\gamma$, and $k_{11}$ all play a role in tumour burden, and this role is differs between the normal development and MYCN amplified neuroblastoma. Therefore, we take each of the neuroblastoma parameter sets corresponding to normal development, MYCN amplified, TERT/ALT amplified, and neuroblastoma with mesenchymal features, fix all parameters other than $\mu$, $\gamma$, and $k_{11}$, and vary these three critical parameters along the parameter space to compute the week 52 tumour burden. This yields the heat maps in Figure~\ref{fig: psweeps}. 

As expected, changes in $k_{11}$, $\mu$ or $\gamma$ do not result in a significant tumour burden in the normal development scenario. However, for the three tumour types, the overall tumour burden increases with the bridge cell proliferation rate $k_{11}$. When $k_{11}=10^{-3}$, no combination of $\gamma$ or $\mu$ result in a substantial tumour burden. As $k_{11}$ increases, distinct regions emerge in the $(\gamma, \mu)$ plane that correspond to either tumour escape or control, with these regions becoming most pronounced when $k_{11}=10^{-1}$. Specifically, a high tumour burden occurs when either $\gamma$ (the rate at which cells that have acquired a mutation acquire new mutation), or $\mu$, the rate at which random mutations are acquired, is elevated. However, a larger tumour burden is observed when $\gamma$ is large. Alternatively. both the values of $\gamma$ and $\mu$ need to remain low ($\approx \leq 10^{-2}$) to achieve minimal tumour burden. Put together, necessary conditions for tumour escape are a sufficiently large value of the bridge cell population proliferation rate, $k_{11}$, and a large value of $\gamma$ and/or $\mu$. Alternatively, necessary conditions for tumour regression are a sufficiently small value of $k_{11}$ and small values for both mutation parameters $\mu$ and $\gamma$. Interestingly, the regions of parameter space that result in either tumour escape or regression are similar between different tumour types but different from the normal development dynamics, again lending credence to the idea that normal development dynamics only lead to transient populations that can result in harmful tumour development if allowed to expand.

\subsection{Treatment relapse}
A major challenge in the treatment of neuroblastoma is the risk of relapse after surgical resection, which is believed to be related to tumour population heterogeneity \cite{Cohn09, Maris10, Zeineldin22awry, Gomez22, Brodeur14Regression}. The reappearance of the tumour is believed to arise from the fact that during surgical resection not all malignant cells are removed, and that they can repopulate the original tumour site. In the previous section, we learnt that while normal development dynamics prevent tumour escape, the clinically abnormal development dynamics associated with various forms of neuroblastoma permit tumour escape when the mutation rates $\mu$ and $\gamma$ are located in a given region of the parameter space. In this section, we extend the question to whether development dynamics might have an impact on tumour reappearance after surgical resection of the tumour. 

\begin{figure}
    \centering
    \includegraphics[width=0.8\linewidth]{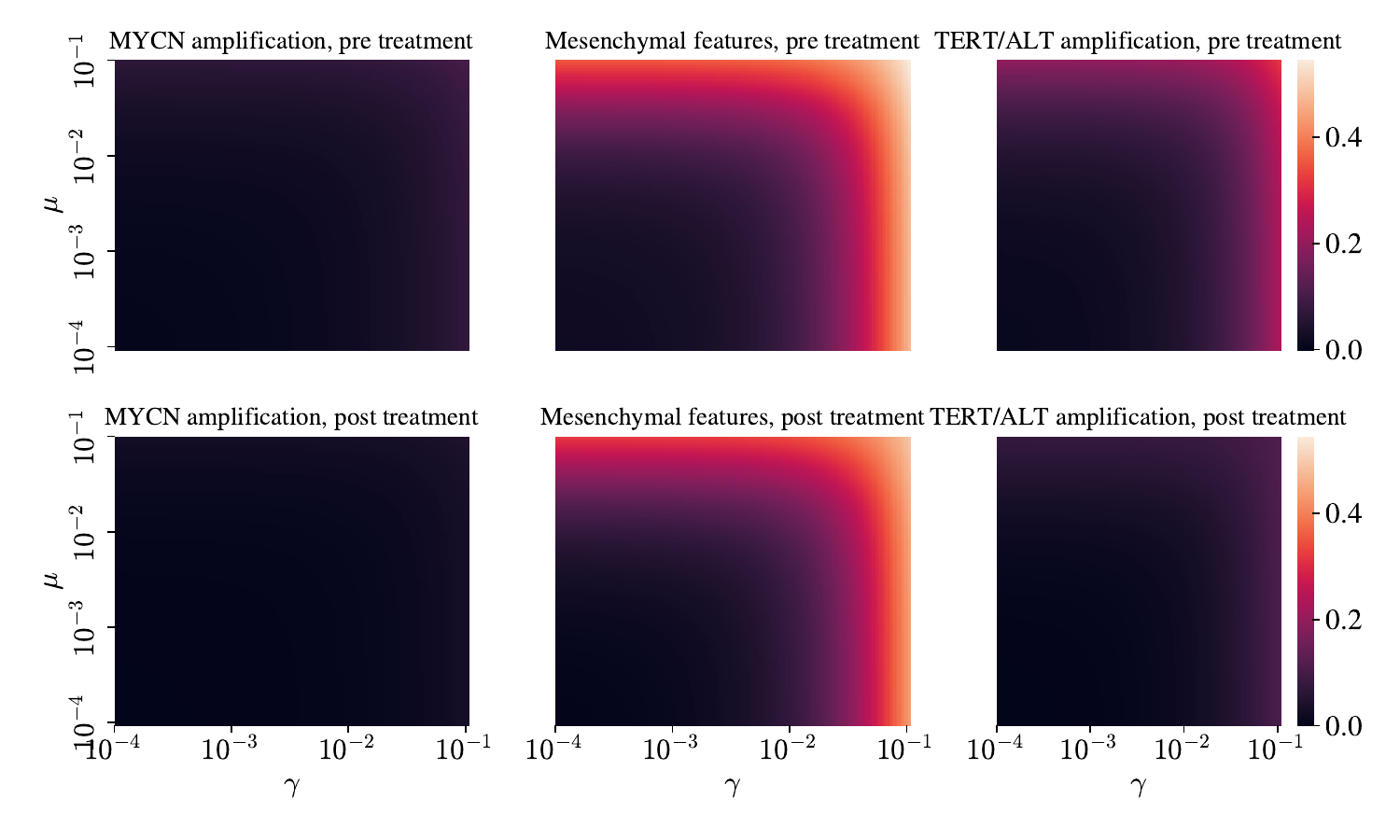}
    \captionsetup{width=1\textwidth}
    \caption{Heatmap showing tumour burden at week 52 pre-treatment and week 104 post-treatment as $\gamma$ and $\mu$ vary for the different tumour types. Surgical resection of the tumour is simulated by decreasing the tumour volume at the end of the 52 week period by 99\%.}
    \label{fig: pre/post treatment}
\end{figure}

\begin{figure}[htp]
    \centering
    \includegraphics[width=1\linewidth]{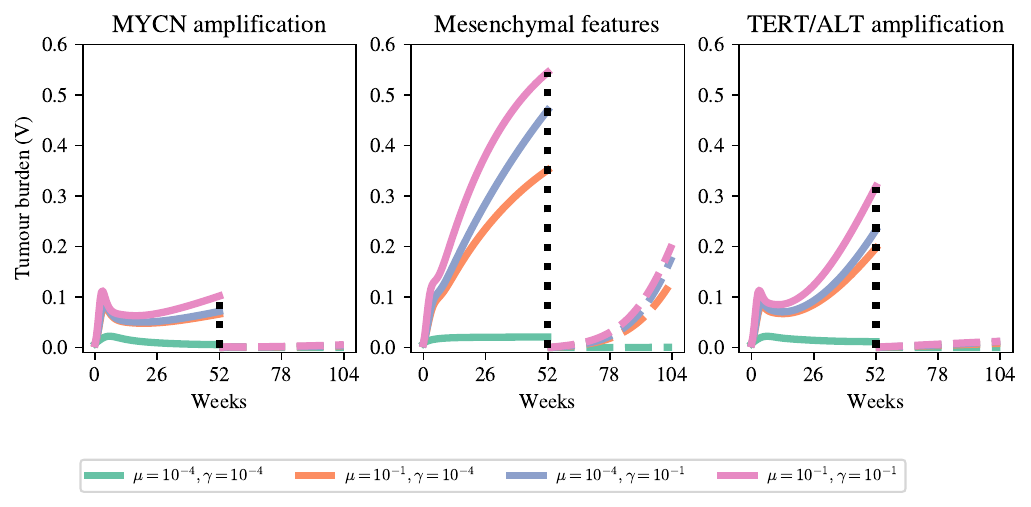}
    \caption{Tumour volume trajectories for the different tumour types as $\gamma$ and $\mu$ vary. Surgical resection is simulated by reducing the tumour volume at week 52 by 99\% as shown by the black dashed line.}
    \label{fig:Treatment}
\end{figure}

To answer this question, we simulated the tumour burden for the three different tumour types defined by the parameters in Table \ref{tab:posteriorMeans} until week 52, then reduced the tumour burden by 99\% by reducing each cell population to 1\% of its size, to mimic surgical resection assuming that the tumour population is spatially mixed. We then continued simulating tumour growth for another 52 weeks to identify the conditions under which relapse occurs. Figure \ref{fig: pre/post treatment} displays the tumour burden at week 52 before treatment and at week 104 after treatment respectively. In Figure \ref{fig: pre/post treatment}, we observe that prior to treatment, the mesenchymal features tumour type has the largest tumour burden compared to the other types, for sufficiently large $\gamma$ and $\mu$, consistent with the predictions from the previous section. Post-treatment, the MYCN amplification and TERT/ALT amplification tumour types respond well to surgical resection, showing no resurgence in tumour burden for any combination of $\gamma$ and $\mu$. However, the mesenchymal features tumour type shows potential for relapse for large $\gamma$ and/or $\mu$. To further tease apart the possible dynamics of treatment relapse in the mesenchymal type neuroblastoma, we compute the tumour burden trajectories across the three neuroblastoma types using different parameter combinations. We show the traces in Figure~\ref{fig:Treatment}. To interrogate to what extent these results depend on the proportion of the tumour removed by resection, we include results corresponding to 95\% and 90\% resection in Appendix~\ref{appendix: additional_trajectories}. These new traces show identical qualitative behaviour to the case of 99\% resection shown in Figure~\ref{fig: pre/post treatment}.
 
The traces in Figure~\ref{fig:Treatment} underscore the potential for relapse in the mesenchymal features tumour type, as tumour volume increases over time when $\gamma$ and/or $\mu$ are sufficiently large, while this does not occur in the other tumour types. This can be related to the fact that the mesenchymal type neuroblastoma tumour contains the largest share of bridge cells out of the different types of tumour (recall the simulations in Figure~\ref{fig:malignancy}). Hence, surgical resection in this tumour type leads to the retention of the largest share of proliferative cells that can develop maligancy, and thus is likeliest to relapse after surgical resection. Additional traces are shown in Appendix \ref{appendix: additional_trajectories} where more of the tumour is left after resection. Unsurprisingly, with a larger tumour burden post treatment, the tumour burden grows larger for all tumour types. Put together, we have found that this model predicts not just that abnormal differentiation dynamics increase the potential of developing sympathoadrenal tissues to form a tumour, but also that treatment outcomes can be different, depending on the differentiation dynamics within the tumour. 

\section{Discussion}
In this work, we have introduced a modelling framework to describe the dynamics of cell differentiation, proliferation, and death in the sympathoadrenal lineage during early embryonic development. The sympathoadrenal lineage contributes to many structures in the embryo, such as the adrenal medulla, and contains the precursor population for neuroblastoma, the most common paediatric extracranial solid cancer. By carefully calibrating our model to clinical data, we found that it can accurately recapitulate the evolution of cell type proportions in the adrenal medulla of healthy patients and those in different categories of neuroblastoma tumours. The calibrated model parameters suggest that key parameters, such as the proliferation rate of immature progenitor cells --  as opposed to that of mature cycling neuroblasts -- and the rate of cell differentiation from a progenitor state to a chromaffin cell state, can explain the vastly different cell type proportions in the different patient categories. These findings contribute to existing efforts in the field to suggest the cell type transitions that are most associated with neuroblastoma development~\cite{Gomez22, Zeineldin22awry}, and can be used as a starting point to understand the genetic programmes and extra-cellular environmental cues responsible for driving abnormal differentiation dynamics leading to neoplasia and eventually neuroblastoma.

To relate abnormal differentiation dynamics during sympathetic nervous system development to cancer, we introduced a phenomenological and phenotypically structured model to describe how malignant traits accumulate in the developing cell population. Taking a moments-based approach, we derived a model to describe the average malignancy within each of the constituent cell populations, and used this to describe the evolution of malignant traits in the population during sympathoadrenal development. Strikingly, we found that, regardless of the speed at which cells with malignant traits acquire subsequent traits, or the rate at which cells randomly lose and acquire malignant traits, healthy differentiation dynamics led to control of malignant traits. On the other hand, each of the cell differentiation programmes that was calibrated for the neuroblastoma categories led to an increase of malignant traits, and in some parameter regimes, to tumour progression. Finally, the model proposed in this paper predicts not just that abnormal differentiation dynamics increase the potential of developing sympathoadrenal tissues to form a tumour, but also that treatment outcomes can be different, depending on the differentiation dynamics within the tumour. The findings arising from this model complement existing theories that the bridge cell population is a potential reservoir for malignant cells~\cite{Jansky21Origin}, and confirm the potential for possible treatments in inducing cell differentiation, such as retinoic acic (RA) in conjunction with other therapies for neuroblastoma~\cite{Matthay99, Korber23, Zeineldin22awry}.

Being a simplified representation of sympathetic nervous system development, our model can be further analysed and extended in several ways. First, one of the key complications arising from neuroblastoma is the potential of the tumour to form metastases~\cite{Cohn09, Maris10, Zeineldin22awry, Gomez22}. While not captured in our model, the ability of the tumour to establish metastases greatly increases the risk of mortality due to neuroblastoma. For this reason, our model can be used as a starting point for modelling efforts aimed to describe the process of neuroblastoma metastasis, or that of a developmental cancer, more generally. This can be done in several ways, although the most immediate extension of this work would be to include a spatial component describing the extra-tumoural environment. This would entail partial differential equation (PDE) models that can carefully recapitulate the ability of each cell type to proliferate, differentiate, and invade the extracellular space. By including further clinically relevant features, such as sources of potential chemoattractants, and angiogenesis, such a model can provide key insights into the cell-intrinsic and cell-extrinsic factors driving neuroblastoma development and progression. 

Furthermore, a central theme in current clinical neuroblastoma research is the ample availability of genetic sequencing data of primary neuroblastoma tumours~\cite{Jansky21Origin, Gomez22, Zeineldin22awry}. With the advent of spatial transcriptomics, one might use various thin tissue slices to reconstruct a three-dimensional distribution of cell types. Further, one of the novelties of calibrating our model to clinical data of neuroblastoma tumours was that, in the absence of time-resolved data, we identified steady states of the model that could be related to the cell type proportions found in neuroblastoma tumours. Having time-resolved patient data would increase the ability to calibrate existing models from data, and potentially learn new models from data~\cite{martinaperez21buq}. However, practically speaking, these data are often lacking since patient biopsies are not often performed during the course of treatment. Rather,  emerging three-dimensional human neuroblastoma patient-derived organoid models may offer a means to obtain time-resolved genomic data. Doing so would contribute to our understanding of the nature of cell type conversions in neuroblastoma, which, despite advances in scRNA sequencing, remain elusive in the context of treatment~\cite{Zeineldin22awry}. Time-resolved patient data could also be used to draw closer connections between mathematical modelling and its use in clinical practice and individualised medicine. For example, if one could link patient characteristics, such as genetic signatures, age of diagnosis, and anatomical location of the tumour to model parameters, modelling approaches could aid as a prognostic tool in individual cancer therapy. Such progress will necessitate not only vast amounts of time-resolved patient data, but also careful model calibration and uncertainty quantification. 

\newpage
\subsubsection*{Authors' contributions}
S.M.-P. conceived the project. S.M.-P. developed the models in this paper and wrote the computational scripts for simulation. S.M.P., R.E.B., J.C.K. and P.M.K. decided the clinical applications of the model to neuroblastoma. L.A.H. carried out the sensitivity analyses and performed the corresponding simulations and figures. S.M.-P. produced the figures and wrote the article, on which all other authors commented. All authors agreed to the publication of this manuscript.

\subsubsection*{Competing interests}
We declare we have no competing interests.

\subsubsection*{Acknowledgements}
S.M.P. would like to thank the Foulkes Foundation for funding. This work was supported by a grant from the Simons Foundation (MP-SIP-00001828, R.E.B.).

\newpage
\bibliography{references}
\bibliographystyle{unsrt}

\newpage
\appendix
\setcounter{equation}{0}\renewcommand\theequation{A\arabic{equation}}
\section{Posterior distributions for model parameters fit to normal adrenal medulla data}
\label{appendix:posterior}
\begin{figure}[htb]
    \centering
    \includegraphics[width=\linewidth]{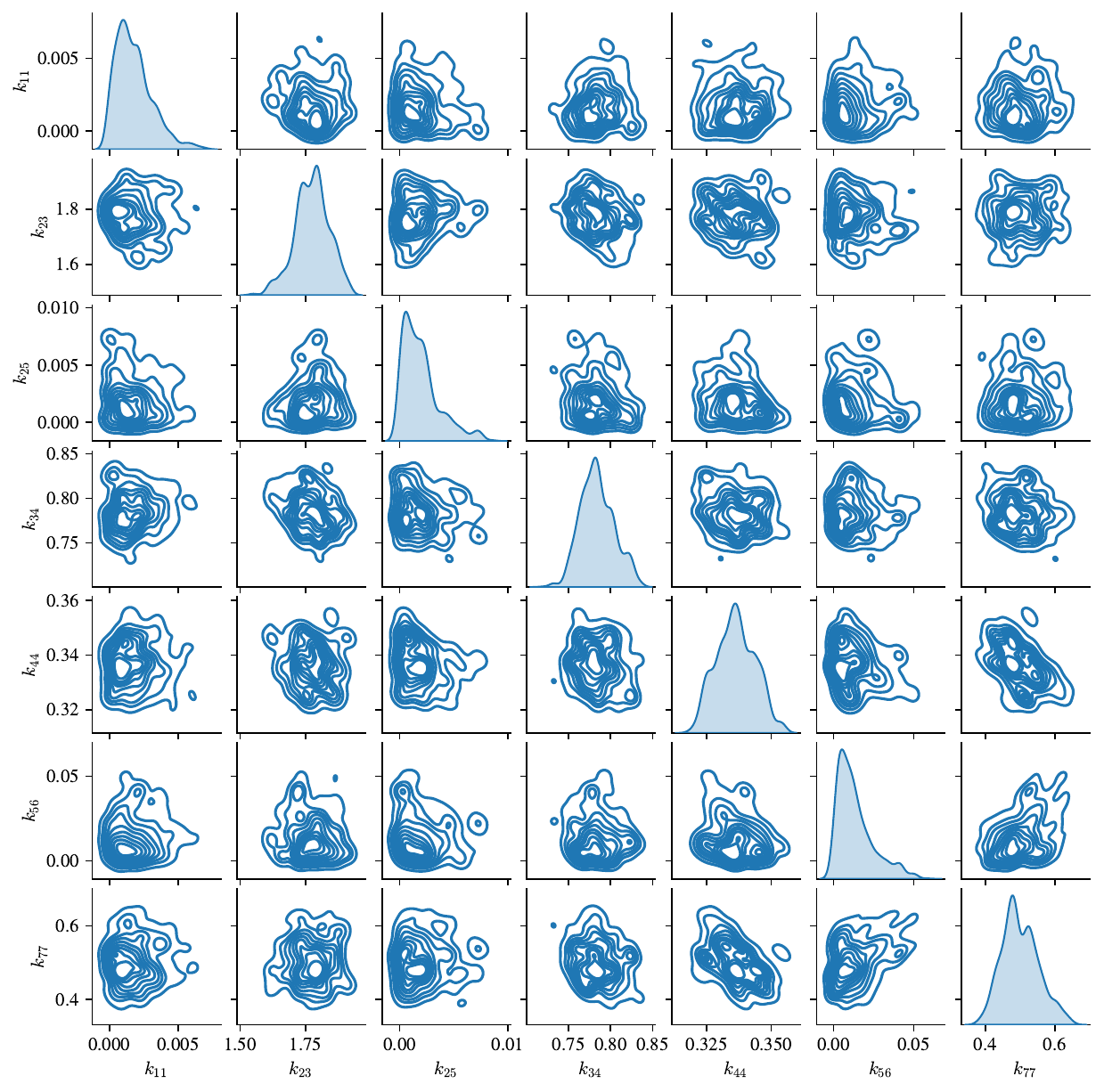}
    \caption{Posterior distribution over model parameters for the model in Equation~\eqref{eq:proportionODE} fit to data of healthy adrenal medulla development.}
    \label{fig:NormalDevelopmentPosteriors}
\end{figure}
\clearpage
\section{ODE derivation for the expected number of malignant cells}
\label{section:MiDerivation}
Multiplying the Fokker-Plank equation in Equation~\eqref{eq:basicFokkerPlanck} and integrating over $m$ yields:
\begin{equation}
    \label{appeq:MEq1}
    \frac{\dd M_j}{\dd t} = \lambda_i M_i + \sum_{j=1}^n k_{ij}M_j + \int_0^1 m\frac{\partial}{\partial m} (b(m) p_i(m,t))+ m\frac{\partial}{\partial m}\left(\sigma(m)\frac{\partial p_i(m,t)}{\partial m}\right)\dd m.
\end{equation}
Note that 
\begin{align*} 
    \int_0^1 m\frac{\partial}{\partial m} (b(m) p_i(m,t)) \dd m = mb(m)p_i(m,t)\vert_{m=0}^{m=1} -\int_0^1 b(m)p_i(m,t)\dd m.
\end{align*}
Given that we chose $b(m) = \gamma (1-m)$ with $\gamma > 0$, the integral becomes
\begin{equation}
    \label{appeq:IBP2}
    \int_0^1 b(m)p_i(m,t)\dd m = \gamma (T_i-M_i).
\end{equation}
Similarly, 
\begin{align*}
    \int_0^1 m\frac{\partial}{\partial m}\left(\sigma(m)\frac{\partial p_i(m,t)}{\partial m}\right)\dd m &= \left.m\sigma(m)\frac{\partial p_i(m,t)}{\partial m}  \right\vert_{m=0}^{m=1} - \int_0^1 \sigma(m)\frac{\partial p_i(m,t)}{\partial m}\dd m\\
    &= -\left( \left.\sigma(m) p_i(m,t)\right\vert_{m=0}^{m=1} - \int_0^1 \sigma'(m) p_i(m,t) \dd m\right).
\end{align*}
Since $\sigma(m) = \mu m(1-m)$, with $\mu > 0$, the above expression integrates to
\begin{equation}
    \label{appeq:IBP1}
    \int_0^1 \sigma'(m) p_i(m,t) \dd m = \mu \int_0^1 (1-2m)p_i(m,t) \dd m = \mu(T_i - 2Mi).
\end{equation}
Substituting Equations~\eqref{appeq:IBP2} and~\eqref{appeq:IBP1} into Equation~\eqref{appeq:MEq1} yields
\begin{equation}
    \frac{\dd M_j}{\dd t} = (\lambda_i-\gamma - 2\mu) M_i +(\gamma+\mu)T_i  + \sum_{j=1}^n k_{ij}M_j.
\end{equation}

\newpage
\section{ODE derivation for the average malignancy}
\label{appendix:malignancy}
Define the mean malignancy, $\mathcal{M}_i$ as $\mathcal{M}_i = M_i/T_i$. We obtain
\begin{align*}
    \frac{\dd \mathcal{M}_i}{\dd t} &= \frac{T_i\cdot\left((\lambda_i-\gamma - 2\mu) M_i +(\gamma+\mu)T_i  + \sum_{j=1}^n k_{ij}M_j\right) - M_i\cdot\left(\lambda_i T_i + \sum_{j=1}^n k_{ij}T_j \right)}{T_i^2}\\
    &= (\gamma + \mu) - (\gamma + 2\mu)\mathcal{M}_i + \frac{T_i\sum_{j=1}^n k_{ij}M_j - M_i\sum_{j=1}^n k_{ij}T_j }{T_i^2} \\
    &= (\gamma + \mu) - (\gamma + 2\mu)\mathcal{M}_i + \frac{\sum_{j=1}^n k_{ij}\mathcal{M}_jT_j - \mathcal{M}_i\sum_{j=1}^n k_{ij}T_j }{T_i}.
\end{align*}
Rearranging,
\begin{equation*}
    \frac{\dd \mathcal{M}_i}{\dd t} = (\gamma + \mu) - (\gamma + 2\mu)\mathcal{M}_i + \sum_{j=1}^n \frac{T_j}{T_i} k_{ij} (\mathcal{M}_j - \mathcal{M}_i),
\end{equation*}
and noting that $T_j/T_i = R_j/R_i$, we finally obtain
\begin{equation}
    \frac{\dd \mathcal{M}_i}{\dd t} = (\gamma + \mu) - (\gamma + 2\mu)\mathcal{M}_i + \sum_{j=1}^n \frac{R_j}{R_i} k_{ij} (\mathcal{M}_j - \mathcal{M}_i).
\end{equation}

\clearpage\section{Additional tumour volume trajectories}
\label{appendix: additional_trajectories}
\begin{figure}[htp]\captionsetup[subfigure]{font=normal}
\captionsetup{width=1\textwidth}
\begin{tabular}{cccc}
\subfloat[95\% reduction]{\includegraphics[width=0.8\linewidth]{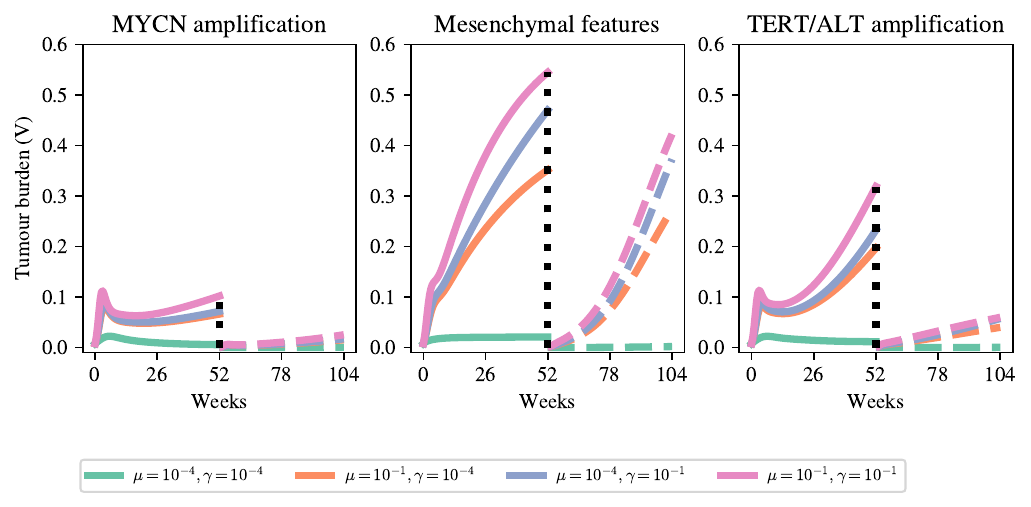}\label{fig: 95 reduction}}\\
\subfloat[90\% reduction]{\includegraphics[width=0.8\linewidth]{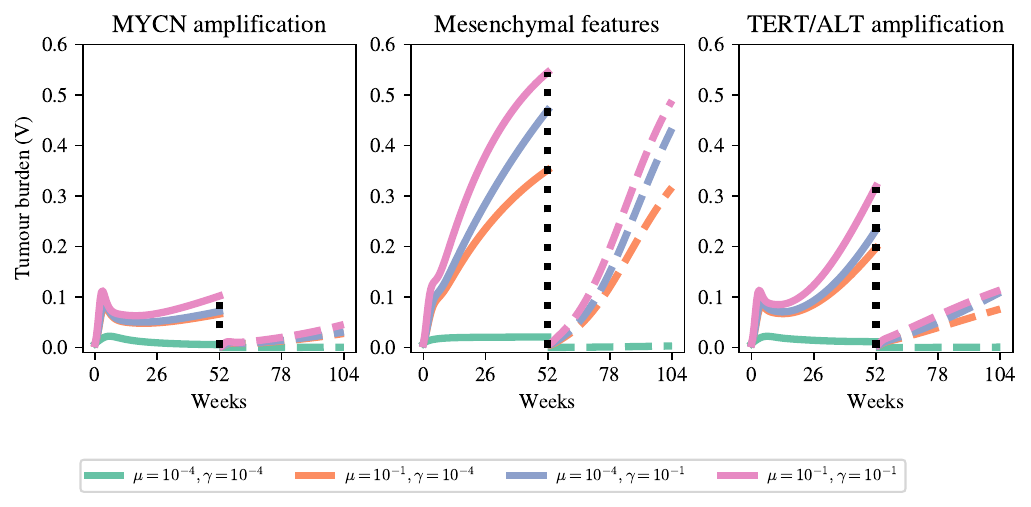}\label{fig: 90 reduction}} \\
\end{tabular}
\centering
\caption{Tumour volume trajectories for the different tumour types as $\gamma$ and $\mu$ vary. Surgical resection is simulated by reducing the tumour volume at week 52 by 95\% in (a) and 90\% in (b) as shown by the black dashed line.}
\label{fig: additional trajectories}
\end{figure}

\end{document}